\documentclass[%
 aip,
 jmp,%
 amsmath,amssymb,
 reprint,%
]{revtex4-1}
\usepackage{graphicx}
\usepackage{dcolumn}
\usepackage{bm}

\begin{document}

\title{The pumpistor: a linearized model of a flux-pumped SQUID for use as a negative-resistance parametric amplifier}

\author{K.M. Sundqvist}
\email{kyle.sundqvist@gmail.com}
\affiliation{Department of Microtechnology and Nanoscience, Chalmers University of Technology, SE-412 96 G\"{o}teborg, Sweden }
\author{S. Kinta\c{s}}
\affiliation{Department of Microtechnology and Nanoscience, Chalmers University of Technology, SE-412 96 G\"{o}teborg, Sweden }
\author{M. Simoen}
\affiliation{Department of Microtechnology and Nanoscience, Chalmers University of Technology, SE-412 96 G\"{o}teborg, Sweden }
\author{P. Krantz}
\affiliation{Department of Microtechnology and Nanoscience, Chalmers University of Technology, SE-412 96 G\"{o}teborg, Sweden }
\author{M. Sandberg}
\affiliation{National Institute of Standards and Technology, 325 Broadway, Boulder, Colorado, 80305, USA}
\author{C.M. Wilson}
\affiliation{Institute for Quantum Computing and Department of Electrical and Computer Engineering, University of Waterloo, Waterloo, Ontario, N2L 3G1, Canada}
\author{P. Delsing}
\affiliation{Department of Microtechnology and Nanoscience, Chalmers University of Technology, SE-412 96 G\"{o}teborg, Sweden }

\date{\today}

\begin{abstract}
We describe a circuit model for a flux-driven SQUID. This is useful for developing insight into how these devices perform as active elements in parametric amplifiers. The key concept is that frequency mixing in a flux-pumped SQUID allows for the appearance of an effective negative resistance. In the three-wave, degenerate case treated here, a negative resistance appears only over a certain range of allowed input signal phase. This model readily lends itself to testable predictions of more complicated circuits.  
\end{abstract}

\maketitle


Parametric amplifiers based on superconducting circuits, though not new\cite{Wahlsten:1978, Yurke:1989}, have recently become the subject of renewed interest \cite{Yamamoto:2008, Castellanos:2007, Wilson:2012, Bergeal:2010,  Hatridge:2011, HoEom:2012} as it is now possible to readily fabricate superconducting circuits which may sustain and amplify coherent states of microwaves close to the quantum limit. Such amplifiers have recently been used for qubit readout\cite{Abdo:2011}, vacuum squeezing\cite{Castellanos:2008}, quantum feedback\cite{Vijay:2012}, and entanglement of propagating microwave photons\cite{ Eichler:2012}. Systems closely related to these amplifiers are also providing new physics, such as photon measurements of fast-tunable resonators\cite{Sandberg:2008} and the observation of the dynamical Casimir effect\cite{Wilson:2011}.

Since topics related to the manipulation of coherent states of light have traditionally been associated with quantum optics, a quantum-optics formalism dominates the commonly encountered explanations of these systems.  However, under suitably small-signal limits, a truly nonlinear reactance may be modeled simply as a \emph{time-varying} reactance.  Under this approximation, the principle of superposition holds and we have the standard lexicon of linear analytical techniques available to us, such as Fourier analysis. In fact, ``classical'' parametric amplifiers were often treated in this linearized manner in literature \cite{Blackwell:1961,Decroly:1973, Howson:1970} generated during the 1960s and 70s.  While this literature was commonly depicting circuits utilizing varactor diodes as active elements, it remains a general premise that a parametrically driven nonlinear reactance leads to frequency mixing.

In this work, a linearized method of analysis allows us to depict the effects of amplification using simple, intuitive models of equivalent electrical circuit elements. We examine the case of a three-wave degenerate parametric amplifier based on a dc Superconducting QUantum Interference Device (SQUID).   Although outside the scope of this work it is also possible to consider the \emph{nondegenerate} case, were an idler tone is introduced and considered separately from a signal tone.  

The parametric interaction is supplied by the SQUID, acting as a tunable, nonlinear inductance.  By way of a mutual inductance to a control line, a time-varying magnetic flux, $\Phi_{\rm{ac}}$, is applied to the SQUID and acts as our pump.  

We will show how the application of a dc and an ac pump flux allows us to treat the SQUID electrically as the well-known Josephson inductance, in parallel to a special circuit element which we introduce as ``the pumpistor.''  We find that the pumpistor defined under these conditions leads to a \emph{phase sensitive} impedance, where the phase angle between the pump and signal tones becomes important. In particular the pumpistor can act as a negative resistance, producing gain. Thus, our treatment presents a simple, analytical, albeit classical understanding of the phase sensitivity associated with degenerate parametric amplification.

Our circuit model of a flux-pumped SQUID allows us to analyze much more complicated circuits in a straightforward way. We use this circuit model to analyze actual experiments performed with a reflection amplifier consisting of a flux-pumped SQUID terminating a $\lambda/4$ transmission-line cavity. We find expressions for the phase-dependent gain, showing how this system operates quite intuitively when analyzed as a negative resistance amplifier. For the quantum engineer, our analysis should prove useful when considering the interface of a flux-pumped SQUID to other RF circuit components, in any number of future novel circuits. 


We start by reviewing the relations between external magnetic flux, effective junction phase, and the current through an ideal SQUID.  We consider this SQUID to be composed of two identical, parallel Josephson junctions, forming a loop which is pumped by an external magnetic flux, $\Phi_{p}$. We neglect capacitive effects in the SQUID, as we assume that all frequencies of interest are well below the Josephson plasma frequency.  The net supercurrent through the SQUID can be expressed as  
\begin{align}
I = \underbrace {{I_c}\left| {\cos \left[ {\pi \;{\Phi _{p}}(t)/{\Phi _0}} \right]} \right|}_{{\rm{``flux" ~term}}}\underbrace {\sin \left[ {\phi (t)} \right]}_{{\rm{``phase"~ term}}}
\label{eqn:SuperCurrent}
\end{align}
where $\Phi_0 = h/(2e)$ is the flux quantum, and $I_c /2$ is the critical current of each of the two Josephson junctions. We identify the current as the product of what we call the ``flux'' term and the ``phase'' term. The phase term contains $\phi(t)$, the superconducting phase difference related to the voltage drop across the SQUID, 
$ V(t) = \left( {\frac{{\Phi _0 }}
{{2\pi }}} \right)\frac{{d\phi(t) }}
{{dt}}
$.


The product of these two terms lead to frequency mixing.  We analyze this by small-signal series expansions for the flux term and the phase term of Eq. (\ref{eqn:SuperCurrent}) which depend on the pump and signal frequencies, respectively. We multiply these expansions to approximate the supercurrent of Eq. (\ref{eqn:SuperCurrent}), and can then define a circuit impedance appropriate for consideration at the angular signal frequency, $\omega_s$.  

We consider the external magnetic flux to be applied with a dc bias and a small ac contribution, as 
\begin{align}
\Phi_{p}(t) = \Phi_{\rm{dc}} + \Phi_{\text{ac}}~\text{cos}(\omega_p t + \theta_{p})
\label{eqn:FluxPerturbation}
\end{align}
where $\omega_p$ is the angular pump frequency, and $\theta_{p}$ its phase angle.  We substitute Eq. (\ref{eqn:FluxPerturbation}) into the ``flux" term of Eq. (\ref{eqn:SuperCurrent}) and series expand about $\Phi_{\rm{dc}}$.  Defining the normalized flux quantities $ F = \pi ~{\Phi _{{\rm{dc}}}}/{\Phi _0}$ and $ \delta f= \pi ~{\Phi _{{\rm{ac}}}}/{\Phi _0}$  gives
\begin{align}
\begin{array}{l}
{{I_c}\cos \left[ {\pi \,{\Phi _{{p}}}(t)/{\Phi _0}} \right]}\\
{\quad \quad  \approx {I_c}\cos \left( F \right) - {I_c}\sin \left( F \right)\,\delta f\cos \left( {{\omega _p}t + {\theta _p}} \right)}.
\end{array}
\label{eqn:IcCosExpand}
\end{align}
We assume the Josephson phase takes the form $\phi(t) = {\phi _s}\cos ( \omega _s t + \theta _s)$.  To treat the full phase term of Eq. (\ref{eqn:SuperCurrent}), we take a Fourier-Bessel expansion where $J_n$ is the $n$-th order Bessel function of the first kind.   
\begin{align}
\sin \left[ {\phi \left( t \right)} \right] = \sum\limits_{n =  - \infty }^\infty  {{J_n}\left( {{\phi _s}} \right)\sin \left[ {n\left( {{\omega _s}t + {\theta _s} + {\textstyle{\pi  \over 2}}} \right)} \right]} 
\label{eqn:PhaseTerm}
\end{align}

We now approximate Eq. (\ref{eqn:SuperCurrent}) by multiplying Eq. (\ref{eqn:IcCosExpand}) and (\ref{eqn:PhaseTerm}).  We find that the resulting supercurrent contains frequency-mixed terms between the pump and signal frequencies.  We choose to retain only terms at the signal frequency and neglect the other mixing products, while recalling that $\omega_p = 2 ~\omega_s$ for the three-wave degenerate case.  This signal current we call $I_{{\omega_s}}(t)$, and the corresponding voltage is $V_{{\omega_s}}(t)$.  We can now define an electrical input impedance at the signal frequency, 
\begin{align}
{Z_{\rm{SQ}}} = \frac{{{V_{{\omega _s}}}\left( t \right)}}{{{I_{{\omega _s}}}\left( t \right)}} = j{\omega _s}{L_{\rm{SQ}}}
\label{eqn:loadImpedance}
 \end{align}
where 
$
L_{{\rm{SQ}}}^{ - 1} = L_J^{ - 1} + L_P^{ - 1}
$, using
\begin{align}
\label{eqn:LJ}   {{L_J}}& = {\frac{{{L_{J0}}}}{{\cos \left( F \right)}}\left[ {\frac{{{\phi _s}}}{{2{J_1}\left( {{\phi _s}} \right)}}} \right]}\\ 
\label{eqn:LP}  {{L_P}}& = {\frac{{ - 2\,{e^{j\Delta \theta }}}}{{\delta f}}\frac{{{L_{J0}}}}{{\sin \left( F \right)}}\left[ {\frac{{{\phi _s}}}{{2{J_1}\left( {{\phi _s}} \right) - 2{e^{j2\Delta \theta }}{J_3}\left( {{\phi _s}} \right)}}} \right]}
\end{align}
with $L_{J0} = \hbar/(2 e I_c)$ and $ \Delta \theta = 2{\theta _s} - {\theta _p}$.

\begin{figure}[t!] 
   \centering
   \includegraphics[width=3.37 in]{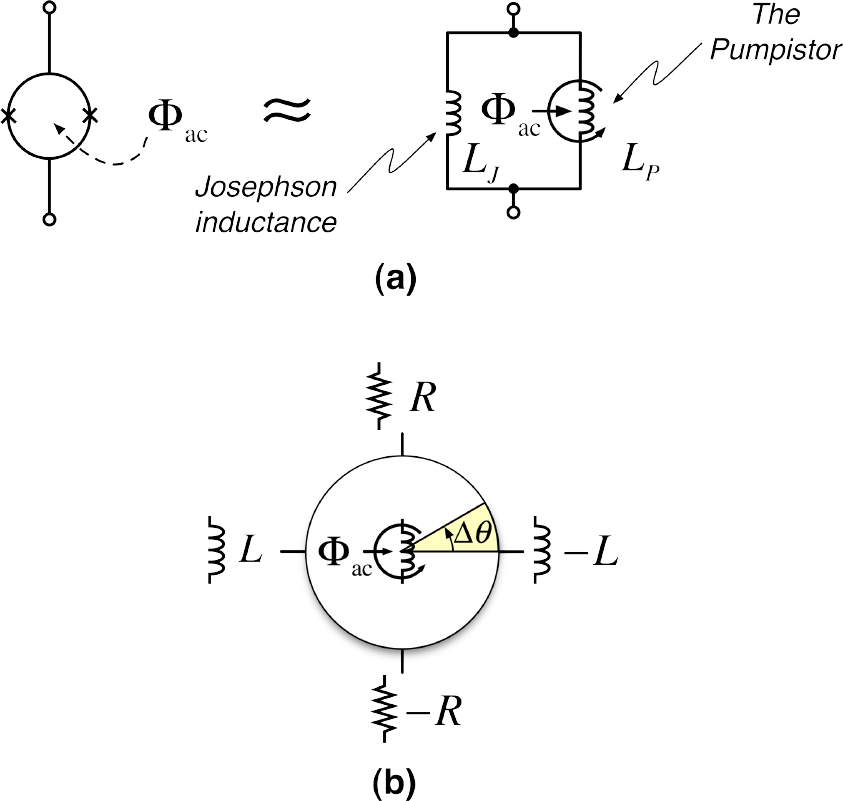} 
   \caption{
 {\bf{(a)}} The flux-pumped SQUID in the small-signal limit can be treated as the quiescent Josephson inductance in parallel to this special element, ``the pumpistor.''   {\bf{(b)}} As a function of phase angle $\Delta \theta$ the pumpistor inductance changes in the complex plane, thereby taking on characteristics of other impedances which introduce a phase sensitivity into the system. 
   }
   \label{fig:pumpsitor_and_circle}
\end{figure}

Thus, we have an equivalent circuit for the driven SQUID (Fig. \ref{fig:pumpsitor_and_circle}a) appearing as the Josephson inductance, $L_J$, in parallel to a new effective element which we call the ``pumpistor," $L_P$.  The pumpistor is defined as an inductance as its impedance is proportional to $j \omega_s$. 

We now examine Eq. (\ref{eqn:LP}) to understand the behavior of the pumpistor.  We see that the pumpistor has a dependence on the applied ac flux amplitude, such that if the flux is turned off the pumpistor impedance becomes infinite (``an open''), and the SQUID impedance appears as the standard Josephson inductance.  Most importantly, the pumpistor inductance also depends on the \emph{phase angle}, $\Delta \theta$.  This introduces the \emph{phase sensitivity} into the circuit (Fig. \ref{fig:pumpsitor_and_circle}b).  When $\Delta \theta = 0$, the pumpistor acts as a negative inductance.\cite{foot:APLpumpsitor1}  Since the two inductances are in parallel, a typically large ($\left| {{L_P}} \right| \gg \left| {{L_J}} \right|$) and negative pumpistor inductance results in a net inductance slightly larger than the Josephson inductance.  At  $\Delta \theta = \pi /2$, the pumpistor inductance becomes imaginary so that its impedance acts as a positive, real resistance adding dissipation.  At  $\Delta \theta = \pi $, the pumpistor has the impedance of a positive inductance. Finally, at  $\Delta \theta = -\pi/2 $,  we find that we have a circuit element that provides a \emph{negative resistance}.  In this state, power is extracted from the pump such that the pumpistor can actively inject power into the external circuit at the signal frequency, providing gain.  We note that the general behavior of a negative resistance used as a reflection amplifier was recently revisited\cite{Clerk:2010}.

Regarding saturation of the ac flux, if we had expanded Eq. (\ref{eqn:IcCosExpand}) to higher powers of $\delta f$ we would have found an equivalent impedance represented by $L_{\rm{SQ}}$ in parallel to more inductive elements.  These extra elements would represent higher-order corrections due to a large ac flux, which account for saturation limitations.  For the specific amplifier discussed later, the ac flux amplitude is small $(\delta f \ll 1)$ and these terms are not necessary. 

Regarding saturation of the Josephson phase, we note that the final, bracketed terms of both Eq. (\ref{eqn:LJ}) and (\ref{eqn:LP}) depict phenomena that would be explained by the ``Duffing nonlinearity"\cite{Kovacic:2011,Wiesenfeld:1984} within the formalism of nonlinear differential equations.  These bracketed terms tend to unity for sufficiently small values of $\phi_s$.  Conversely, larger values of $\phi_s$ will affect the gain and bandwidth response expected from the amplifier.  Although these effects are typically associated with saturation due to $\phi_s$, \emph{bifurcation} may also arise where multiple solutions of $\phi_s$ are allowed in the amplifier circuit.  In this work, we present a treatment of our amplifier considering only the limit of a small $\phi_s$.

Consider, generally, that the pumpistor should act as the active element in some amplifier topology.  As the pumpistor is always found in parallel to the smaller, passive inductance $L_J$, it is heavily shunted and its effect suppressed.  In a useful device, the impedance of $L_J$ should therefore be nulled by implementing a system containing other reactances.  

\begin{figure}[t] 
   \centering
   \includegraphics[width= 3.37 in ]{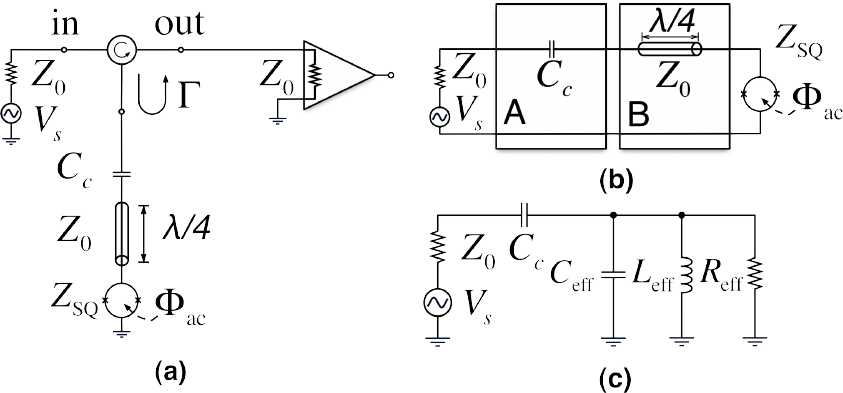}   
   \caption{
   {\bf{(a)}} The pumped cavity reflectometry amplifier showing physical ports,  {\bf{(b)}} as a two-port network,  {\bf{(c)}} equivalent lumped circuit
   }
   \label{fig:cavity_for_APL}
\end{figure}

A parametric amplifier topology commonly used by multiple groups, including our own, is a flux-pumped SQUID terminating a superconducting quarter-wave transmission line resonator which is coupled to a transmission line via a capacitor. As shown in Fig. (\ref{fig:cavity_for_APL}a), a signal incident at the input enters a circulator and is reflected by the amplifier with reflection coefficient $\Gamma$ to the output port. The output power is ${P_{\rm{out}}} = {\left| \Gamma  \right|^2}  P_{\rm{in}} $, allowing us to define the power gain as $ G = P_{\rm{out}} / P_{\rm{in}}= |\Gamma|^2$.  The negative resistance of the pumpistor allows $|\Gamma|$ to become greater than unity, thus achieving gain.  

To find the expression for gain, we solve the system for the reflection coefficient.  We use S-parameters for two-port networks as in Fig. (\ref{fig:cavity_for_APL}b) to find the system's reflection coefficient.  We consider the coupling capacitor, $C_c$, to have the S-matrix $S_A$. This is connected to a matched ($Z_0 = 50 ~\Omega$) transmission line cavity represented by $S_B$.
\begin{align}
{{S_A}} & = \left( {\begin{array}{*{20}{c}}
{\frac{1}{{1 + j2\omega C_c{Z_0}}}}&{1 - \frac{1}{{1 + j2\omega C_c{Z_0}}}}\\
{1 - \frac{1}{{1 + j2\omega C_c{Z_0}}}}&{\frac{1}{{1 + j2\omega C_c{Z_0}}}}
\end{array}} \right)\\
{{S_B}} & = \left( {\begin{array}{*{20}{c}}
0&{{e^{ - \gamma l}}}\\
{{e^{ - \gamma l}}}&0
\end{array}} \right)
\end{align}
We consider the propagation constant to be $\gamma l = j \omega \sqrt{(Ll)(Cl)} + \alpha l$.  $L$ and $C$ are the inductance and capacitance per unit length, respectively, while $\alpha$ specifies the attenuation constant, and $l$ is the length of the transmission line.  The load reflection is of the standard form
$
{\Gamma _{{\rm{SQ}}}} = \left( {{Z_{{\rm{SQ}}}} - {Z_0}} \right)/\left( {{Z_{{\rm{SQ}}}} + {Z_0}} \right)
$,
where $Z_{\rm{SQ}}$ is from Eq. (\ref{eqn:loadImpedance}). We find the \emph{total} reflection coefficient becomes
\begin{align}
\Gamma  = {S_{A,11}} + \frac{{{S_{A,21}}{S_{A,12}}{\Gamma _{{\rm{SQ}}}}{e^{ - 2\gamma l}}}}{{1 - {S_{A,22}}{\Gamma _{{\rm{SQ}}}}{e^{ - 2\gamma l}}}}.
\label{eqn:TotalReflection}
\end{align}
which, upon taking its squared amplitude, provides the expression for gain.

We can also use Eq. (\ref{eqn:TotalReflection}) to define an input impedance into this entire chain. We use the relation
$
{Z_{in}} = {Z_0}  (1 + \Gamma ) / (1 - \Gamma )
$
to define an input impedance. The transmission line is designed for operation close to its $\lambda/4$ resonance so that $\omega_s \approx \pi/(2 \sqrt{  (Ll) (Cl) }) \approx 2\pi \times 5~\rm{GHz}$.  Using a series expansion in $\omega$ of the cavity admittance ${\left[ {{Z_{in}} - \left( {j\omega {C_c}} \right)} \right]^{ - 1}}$ near the $\lambda/4$ resonance and assuming $C_c \ll Cl$, we can find an equivalent lumped circuit (Fig. \ref{fig:cavity_for_APL}c) using the following relations,
\begin{align}
{{L_{{\rm{eff}}}}}&{ = \frac{{8\,Ll}}{{{\pi ^2} + \frac{{{\pi ^4}}}{{4{{\left( {Ll} \right)}^2}}}\left( {{\rm{Re}}{{[{L_{{\rm{SQ}}}}]}^2} - {\rm{Im}}{{[{L_{{\rm{SQ}}}}]}^2}} \right)}}} 
\end{align}
\begin{align}
\begin{array}{*{20}{l}}
{{C_{{\rm{eff}}}}}&{ = \frac{{Cl}}{2} + \frac{{Cl\;{\rm{Re}}[{L_{{\rm{SQ}}}}]}}{{Ll}} + \frac{{{\pi ^2}Cl\left( {{\rm{Re}}{{[{L_{{\rm{SQ}}}}]}^2} - {\rm{Im}}{{[{L_{{\rm{SQ}}}}]}^2}} \right)}}{{8{{\left( {Ll} \right)}^2}}}}
\end{array}
\end{align}
\begin{align}
{{R_{{\rm{eff}}}}}&{ =  - \frac{{2{Z_0}\left( {Ll} \right)}}{{\pi {\rm{Im}}[{L_{{\rm{SQ}}}}]}}}.
\end{align}

The series resonance frequency is given by ${\omega _{0}} = {\left[ {{L_{{\rm{eff}}}}\left( {{C_c} + {C_{{\rm{eff}}}}} \right)} \right]^{ - 1/2}}$.  Approximating the response near this resonance as harmonic, we find the following internal and external quality factors,
\begin{align}
 \label{eqn:Qint}
{{Q_{{\rm{int}}}}}&{ \approx {R_{{\rm{eff}}}}\left( {{C_{{\rm{eff}}}} + {C_c}} \right){\omega _0}}\\
{{Q_{{\rm{ext}}}}}&{ \approx {{\left( {{Z_0}{\omega _0}C_c^2} \right)}^{ - 1}}\left( {{C_{{\rm{eff}}}} + {C_c}} \right).}
 \label{eqn:Qext}
\end{align}

An approximation for the extremum value of the reflection coefficient at resonance is given by
$
{\Gamma _m} =  ( Q_{{\rm{ext}}} - Q_{{\rm{int}}} )/( Q_{{\rm{ext}}} + Q_{{\rm{int}}}).
$
For conditions with no pumping and where internal dissipation is present we have $Q_{\rm{int}} > 0 $, and $|\Gamma_m|$ has a value somewhere between zero and unity.  For the case where the the flux pump is on and the pumpistor creates a net \emph{negative} load resistance, we find $Q_{\rm{int}} < 0 $. This case represents a net injection of power and $|\Gamma_m| > 1$. The condition for $|\Gamma_m| = \infty$ is where $Q_{\rm{int}} = -Q_{\rm{ext}}$, and depicts the threshold for the amplifier to become a parametric oscillator\cite{Wilson:2010, Wilson:2012, Wustmann:2013, Dykman:1998}. All values of reasonable amplifier gain must occur quite near this threshold, but with $Q_{\rm{int}} < -Q_{\rm{ext}}$.  In this linear model, the region for parametric oscillation ($ -Q_{\rm{ext}} < Q_{\rm{int}} < 0 $) represents instability.   

To construct and bias this device for operation as an amplifier, $Q_{\rm{int}}$ must be only slightly more negative than $-Q_{\rm{ext}}$.  Evaluating the threshold condition $Q_{\rm{int}} = -Q_{\rm{ext}}$ using Eq. (\ref{eqn:Qint}) and (\ref{eqn:Qext}), with both $C_c \ll Cl$ and $|L_{\rm{SQ}}| \ll Ll$, we find the following,
\begin{align}
\left| {{\mathop{\rm Im}\nolimits} [{L_{{\rm{SQ}}}}]} \right| \approx \frac{{\pi C_c^2}}{{2{{\left( {C l } \right)}^2}}}L l.
\end{align}
So external flux and design parameters should be adjusted to approximately meet this criterion.

\begin{figure}[t!] 
   \centering
      \includegraphics[width=3.37 in]{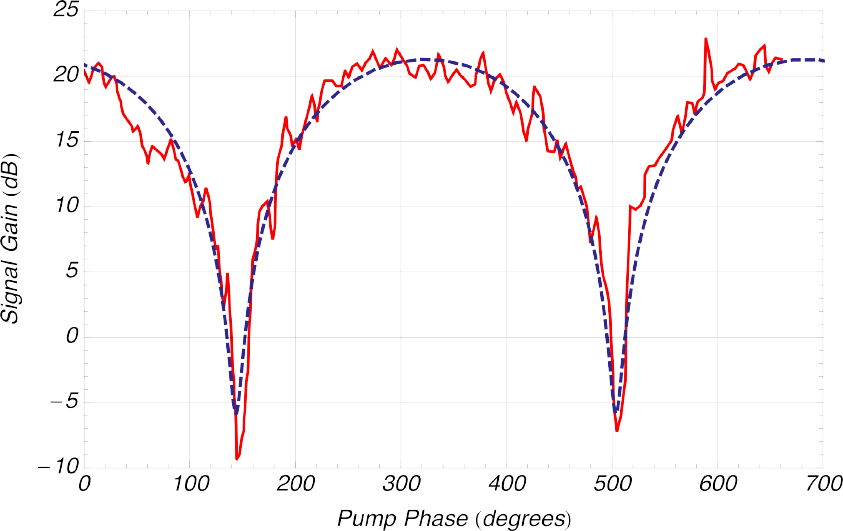} 
   \caption{
Our circuit theory (blue, dashed line) and experiment (red, solid line) for the pumped cavity amplifier. This shows signal gain at a static frequency near resonance, as a function of pump phase.
   }
   \label{fig:Sandberg_phase}
\end{figure} 

As an interesting comparison to data, we examine the response of this amplifier as a function of relative pump phase.  Consider the signal phase angle $\Delta \theta$.  This was defined relative to the signal angle, $\theta_s$, of the \emph{Josephson phase}, $\phi(t)$.  We relate this to the angle of the \emph{SQUID voltage}, $\theta_s^V \equiv  \theta_s + \pi/2$.  Furthermore, we reference $\theta_s^V$ to the angle $\theta_{in}$ of the input's \emph{source voltage}, $V_s$, through the use of a transfer function derived by scattering matrices.   We then map the gain of the amplifier as a function of changing the relative pump phase, $2 \theta_{in} - \theta_{p}$.  This is demonstrated in Fig. \ref{fig:Sandberg_phase} and compared to data.  Experimentally, the physical device was operated at $T\approx20~mK$ and consisted of an Al coplanar waveguide lithographed on a Si substrate, with junctions fabricated using standard two-angle evaporation. The model parameters were found to be: $C_c = 5.4 ~fF$, $Cl = 0.8~pF$, $Ll =2.3~nH $, $I_c =1.7 ~ \mu A $, $\Phi_{\rm{dc}}=0.38~\Phi_0$, $\Phi_{\rm{ac}}=7.8 \times 10^{-4}~\Phi_0$, and $\alpha l = 3.0 \times 10^{-4}$.

In conclusion, we have determined the effective impedance for the parametrically flux-pumped SQUID in the three-wave, degenerate case.  This impedance appears as the standard Josephson inductance in parallel to a new phase-dependent element, which we call the pumpistor.  In addition to a dependence on the ac flux amplitude, the pumpistor has a dependence on the signal-pump phase angle. This results in \emph{phase sensitive} amplification.  When the phase angle is such that the pumpistor has an impedance component which is both real and negative, it is possible to achieve gain.  Within this circuit framework we have analyzed our experimental three-wave degenerate amplifier, thereby validating our phase-dependent model against physical data.


\bibliographystyle{apsrev}

\end{document}